\documentclass[graybox]{svmult}

\usepackage{url}
\usepackage[strings]{underscore} 
\usepackage{mathptmx}       
\usepackage{helvet}         
\usepackage{courier}        
\usepackage{makeidx}         
\usepackage{graphicx}        
\usepackage{multicol}        
\usepackage[bottom]{footmisc}
\usepackage[numbers]{natbib} 
\makeindex

\newcommand{\beq}{\begin{eqnarray}}
\newcommand{\eeq}{\end{eqnarray}}

\newcommand{\be}{\begin{eqnarray*}}
\newcommand{\ee}{\end{eqnarray*}}

\newcommand{\bqa}{\begin{eqnarray}}
\newcommand{\eqa}{\end{eqnarray}}

\newcommand{\bel}[1]{\begin{eqnarray}\label{#1}}
\newcommand{\beal}[1]{\bel{#1}}
\newcommand{\eel}{\end{eqnarray}}
\newcommand{\eea}{\end{eqnarray}} 

\newcommand{\rf}[1]{Eq.~(\ref{#1})}
\newcommand{\rfn}[1]{~(\ref{#1})}

\newcommand{\p}{\partial}

\newcommand{\f}[2]{\frac{#1}{#2}}

\newcommand{\ed}{{\cal E}}       
\newcommand{\sd}{{\cal S}}       
\newcommand{\peq}{{\cal P}}     
\newcommand{\pT}{{\cal P}_T}   
\newcommand{\pL}{{\cal P}_L}   

\def\umu{u^\mu}                          
 
\def\unu{u^\nu}
\def\unul{u_\nu}

\def\pt{p_T}                                  
\def\pl{p_L}                                  

\def\gmunu{g^{\mu\nu}}
\def\Dmunu{\Delta^{\mu\nu}}

\def\Tmunu{T^{\mu \nu}}
\def\Tmumu{T^\mu_{\,\,\,\mu}}
\def\TmunuEQ{T^{\mu\nu}_{\rm EQ}}
\def\Tmunueq{T^{\mu\nu}_{\rm eq}}

\def\Pimunu{{\Pi^{\mu \nu}}}
\def\pimunu{{\pi^{\mu \nu}}}

\def\pinumu{{\pi^{\nu \mu}}}
\def\pimumu{\pi^\mu_{\,\,\,\mu}}

\def\sigmamunu{{\sigma^{\mu \nu}}}

\def\trel{\tau_{\rm rel}}

\begin{document}
\title*{Hydrodynamic description of ultrarelativistic heavy-ion collisions}
\titlerunning{Hydrodynamic description of ultrarelativistic heavy-ion collisions} 
\author{Wojciech Florkowski}
\authorrunning{W.~Florkowski}  
\institute{Wojciech Florkowski \at Institute of Nuclear Physics, Polish Academy of Sciences, PL-31-342 Krak\'ow, Poland,\\ 
Jan Kochanowski University, PL-25-406 Kielce, Poland, and \\
ExtreMe Matter Institute EMMI, GSI, D-64291 Darmstadt, Germany \\
\email{Wojciech.Florkowski@ifj.edu.pl}
} 
\maketitle
\abstract{Recent theoretical developments of relativistic hydrodynamics applied to ultrarelativistic heavy-ion collisions are briefly reviewed. In particular,  the concept of a formal gradient expansion is discussed, which is a tool to compare different hydrodynamic models with underlying microscopic theories.}
%


\section{Introduction}
\label{sec:1}
\sectionmark{Introduction}

The data collected in heavy-ion experiments performed at RHIC and the LHC are interpreted as the
evidence for formation of an equilibrated strongly interacting matter that exhibits collective, fluid-like
behavior. This new state of matter has been named a strongly interacting quark-gluon plasma (QGP). 
Theoretical description of plasma space-time evolution is based on vast applications of
relativistic viscous hydrodynamics whose methods and applicability range are now very
intensively studied. 

In these lectures, several new developments within relativistic hydrodynamics used in the context
of heavy-ion physics are presented. These developments refer, in particular,  to: 1) the very concept of relativistic
hydrodynamics as a universal description of systems approaching local equilibrium, 2) different 
formulations of relativistic hydrodynamics and methods that can be used to make comparisons
between such formulations, and 3) the physical concept of early hydrodynamization that has replaced recently
the idea of early thermalization of matter produced in the collisions. 

The field of relativistic hydrodynamics is very broad and very actively analyzed at the moment, so these
lectures cover necessarily only a few topics. For a general text that tries to answer the question what
relativistic hydrodynamics is we refer to~\cite{Florkowski:2017olj}. Other recent reviews on
relativistic hydrodynamics used for heavy-ion collisions can be found 
in \cite{Romatschke:2009im,Jeon:2015dfa,Jaiswal:2016hex,Alqahtani:2017mhy,Yan:2017ivm}. 
The classical text on hydrodynamics is \cite{LLfluid}
and its modern version can be found in \cite{rezzolla2013relativistic}. Several textbooks are also
available now that discuss relativistic hydrodynamics as a part of heavy-ion physics 
\cite{Yagi:2005yb,Vogt:2007zz,Florkowski:2010zz}. The connections between hydrodynamics,
heavy-ion physics and string theory are discussed in 
\cite{CasalderreySolana:2011us,DeWolfe:2013cua,Heller:2016gbp}.

In the remaining part of Introduction we introduce the standard model of heavy-ion collisions,
discuss basic hydrodynamic concepts, define perfect-fluid and Navier-Stokes hydrodynamics,
and present the insights from the AdS/CFT correspondence and the kinetic theory
in the relaxation time approximation (RTA).

\subsection{Standard model of heavy-ion collisions}
\label{sec:1.1}
\sectionmark{Standard model of heavy-ion collisions}
%
Nowadays, one speaks often about the standard model of ultrarelativistic heavy-ion collisions, which
separates the space-time evolution of the produced matter into three stages. 

The first stage,  lasting for about 1--2 fm/c after the initial impact, describes the matter that is highly 
out of equilibrium.  To large extent, during this early time, the initial conditions for further evolution of 
matter are fixed. They can be determined by using, for example, the Glauber model which is
based on simple geometric concepts~\cite{Florkowski:2010zz}. In this case, the initial energy density of the system
in the transverse plane reflects the distribution of nucleons in the colliding nuclei. 
In the first stage the hard probes are emitted: heavy quarks, jets and energetic photons.  
From a new theoretical perspective, the first stage can be called the hydrodynamization stage, 
i.e., the stage where the produced matter becomes eventually well described by equations of 
viscous hydrodynamics.

The second stage describes the hydrodynamic expansion of matter and lasts for about
10 fm/c (for central collisions of large nuclei). During this stage, a phase transition from the
quark-gluon plasma to a hadronic gas takes place.  It is built into the hydrodynamic equations 
through the use of the appropriate equation of state~\cite{Petreczky:2012rq}. The hydrodynamic expansion leads 
to local equilibration, namely, to the situation where dissipative processes become
negligible. The evidence for local equilibration comes from numerous successes of so
called thermal models which analyze the ratios of hadronic abundances, for example, see~\cite{Floris:2014pta}
and references therein. 

The final, third stage of the evolution of matter is called the freeze-out and describes 
decoupling of strongly interacting gas of hadrons into individual particles that
are finally detected in heavy-ion experiments. In this lecture we shall 
not discuss the freeze-out process which is analyzed in a separate lecture by 
Ryblewski. We shall not discuss the effects of finite baryon density either,
just concentrating on the collisions at the top available energies at RHIC and
the LHC.

\begin{figure}[t] 
\begin{center}
\includegraphics[angle=0,width=0.9\textwidth]{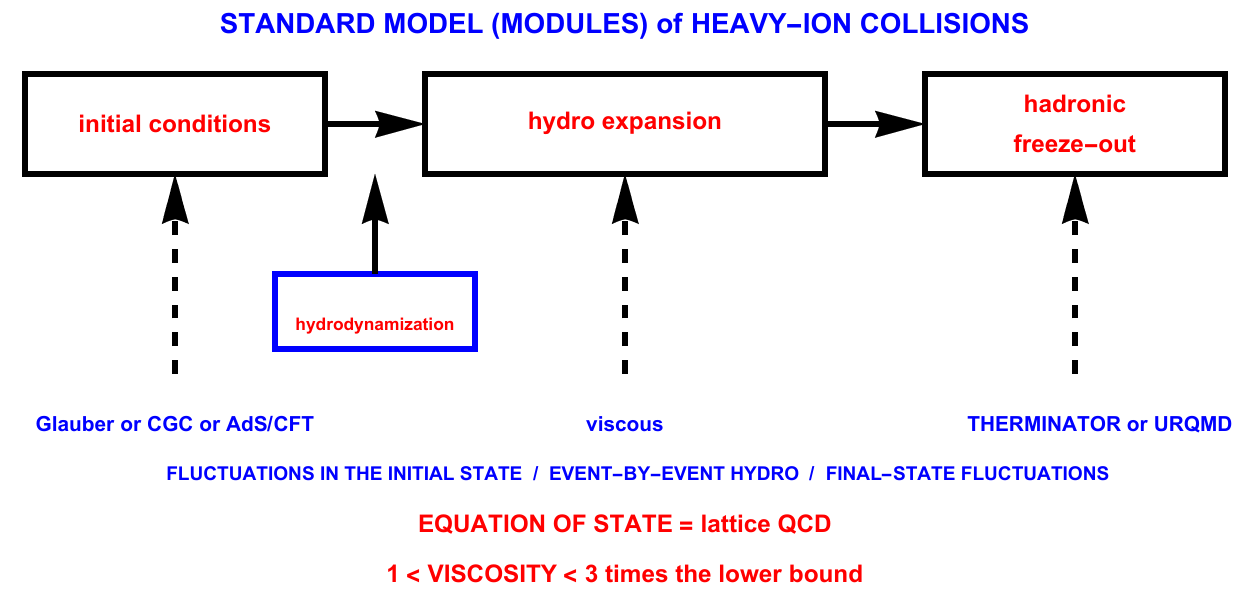}
\end{center} 
 \caption{Schematic view of the standard model of heavy-ion collisions where the evolution of matter
 is separated into three stages (description in the text).}
 \label{fig:SMnew}
\end{figure}

%
\subsection{Basic hydrodynamic concepts}
\label{sec:1.2}
\sectionmark{Basic hydrodynamic concepts}
%
In classical physics hydrodynamics deals with liquids in motion. It is a subdiscipline of fluid mechanics 
(fluid dynamics) which deals with both liquids and gases~\cite{LLfluid}. Liquids, gases, solids and plasmas are states 
of matter, characterized locally by macroscopic quantities, such as energy density, temperature or pressure.
States of matter differ typically by compressibility and rigidity -- liquids are less compressible than gases, 
solids are more rigid than liquids. We learn at school that a typical liquid conforms to the shape of its container 
but retains a (nearly) constant volume independent of pressure. A classical, natural explanation of different 
properties of liquids, gases, solids and plasmas is achieved  within atomic theory of matter. 

It is important for modern applications that hydrodynamics, similarly to thermodynamics, may be formulated 
without explicit reference to microscopic degrees of freedom~\footnote{This point of view will be explicitly
presented below in Sec.~\ref{sec:3.3}.}. This is important if we deal with a strongly interacting QGP
interacting matter --- in this case neither hadronic nor partonic degrees of freedom seem to be adequate 
degrees of freedom.

The information about the state of matter is, to large extent, encoded in the structure of its energy-momentum tensor
(equation of state, kinetic (transport) coefficients). This structure may be a priori determined by modeling 
of heavy-ion collisions. We are in some sense lucky that this scenario has been indeed realized. This is so, 
because the created system indeed evolves toward local equilibrium state, as we pointed out in the 
previous section discussing the applicability of thermal models.

The current understanding is that hydrodynamics can be treated as an effective theory describing approach 
of physical systems toward local equilibrium~\cite{Florkowski:2017olj}. During such an equilibration process different modes in the
system are excited, which can be classified as either transient or long-lived ones. The former are called 
the non-hydrodynamic modes, while the latter are called hydrodynamic ones. Genuine hydrodynamic
behavior is attributed to the hydrodynamic modes, whose lifetime can be made larger by lowering the
spatial variations of various physical variables such as energy density or pressure.

From this perspective it may come as surprise that hydrodynamic models used at the moment to analyze the
heavy-ion data are based on the systems of equations which include both non-hydrodynamic
and hydrodynamic modes. This is so because such models are constructed as approximations to 
microscopic theories which naturally include all kinds of excitations. The hydrodynamic modeling
reveals the authentic hydrodynamic behavior of the systems only if it is not sensitive to the transient 
modes~\cite{Romatschke:2016hle,Spalinski:2016fnj,Romatschke:2017vte}.

%
\subsection{From global to local equilibrium}
\label{sec:1.3}
\sectionmark{From global to local equilibrium and Navier-Stokes hydrodynamics}
%
As hydrodynamical behavior is characteristic for system approaching equilibrium
let us introduce first the concepts of global and local equilibrium which are defined
by the specific forms of the energy-momentum tensor. The {\it global equilibrium} 
energy-momentum tensor in the  fluid rest-frame is given by the expression~\cite{LLfluid}
\bel{equilibrium}
\TmunuEQ = \left(
\begin{array}{cccc}
\ed_{\rm EQ} & 0 & 0 & 0 \\
0 &  \peq(\ed_{\rm EQ}) &  0 &  0\\
0 & 0 &   \peq(\ed_{\rm EQ}) & 0 \\
0 & 0 &  0 &  \peq(\ed_{\rm EQ}) \\
\end{array} \right).
\eel
Here we assume that the equation of state is known, so the equilibrium pressure $\peq$ is 
a given function of the energy density $\ed_{\rm EQ}$. In an arbitrary frame of reference
we have
\begin{eqnarray}
\TmunuEQ = \ed_{\rm EQ} \umu \unu - \peq(\ed_{\rm EQ}) \Delta^{\mu \nu}, 
\label{TmunuEQ}
\end{eqnarray}
where $\umu$ is a constant velocity, and $\Dmunu$ is the operator that
projects on the space orthogonal to $\umu$, namely~\footnote{We use the metric tensor $g_{\mu\nu}$ = diag(+,-,-,-).}
\begin{equation}
\Dmunu = \gmunu-\umu \unu, \quad \Dmunu u_\nu = 0.
\label{defDelta}
\end{equation}
The concept of {\it local equilibrium} is introduced by allowing the
variables $\ed$ and $\umu$ to depend on the spacetime point $x$
\begin{eqnarray}
\Tmunueq(x) = \ed(x) \umu(x) \unu(x) - \peq(\ed(x)) \Dmunu(x).
\label{TmunuPerf} 
\end{eqnarray}
Here, the subscript ``eq'' refers to local thermal equilibrium. 
The energy-momentum tensor~\rfn{TmunuPerf} describes {\it perfect fluid}.
Local effective temperature $T(x)$ is determined by the
condition that the equilibrium energy density at this temperature agrees with
the non-equilibrium value of the energy density, namely
\begin{eqnarray}
\ed_{\rm EQ}(T(x))  = \ed_{\rm eq}(x) = \ed(x).
\label{LM0} 
\end{eqnarray}
The variables $T(x)$ and $\umu(x)$ are fundamental fluid/hydrodynamic variables.
The relativistic perfect-fluid energy-momentum tensor \rfn{TmunuPerf} is the most 
general symmetric tensor which can be expressed in terms of these variables 
without using derivatives.

Dynamics of the perfect fluid is determined by the conservation
equations of the energy-momentum tensor
\bel{conservationeq}
\p_\mu \Tmunueq = 0.
\eel
These are four equations for the four independent hydrodynamic fields ($T(x)$
and three independent components of $\umu(x)$). In this way one obtains 
a self-consistent theoretical framework. It is important to notice that 
dissipation does not appear in the perfect-fluid dynamics. The projection
of \rf{conservationeq} along $u_\nu(x)$ leads to the entropy conservation law
\beq
\p_\mu (\sd \umu) = 0.
\eeq
The quantity $\sd$ is the entropy density. The other three components in \rf{conservationeq} correspond to the non-relativistic
Euler equation.

%
\subsubsection{Landau and Bjorken models}
\label{sec:1.3.1}
\sectionmark{Landau and Bjorken models}
%

The two famous perfect-fluid hydrodynamic models of particle production were formulated in the past 
by Landau~\cite{Landau:1953gs} and Bjorken~\cite{Bjorken:1982qr}. Till now we refer frequently to them 
discussing different initial conditions used for hydrodynamic equations.  In the Landau model~\cite{Landau:1953gs}, 
the matter produced in a collision forms initially 
a highly compressed disk. The equations of perfect fluid are then used to determine a one-dimensional 
expansion of matter along the collision axis. The Landau initial conditions correspond to so called full stopping 
scenario  with high initial baryon number density in the central part of the collision region.
On the other hand, in the Bjorken model we deal with so called transparency regime and typically
negligible baryon number density in the central region. The Bjorken model~\cite{Bjorken:1982qr} is motivated by
the observation that fast particles are produced later and further away from the collision center than the slow ones. 
It is possible to account for this effect in the hydrodynamic  description by imposing special initial conditions. 
In the Bjorken model they are implemented by the assumption that hydrodynamic expansion 
is invariant with respect to longitudinal Lorentz boosts (commonly known as boost invariance). 

A popular expectation is that the Landau model is more appropriate for low-energy collisions, while the Bjorken 
model is suitable for description of high-energy processes. In practice one encounters sometimes Landau-like
features at high energies and Bjorken-type features at low energies. For example, the rapidity distribution of the 
produced pions has usually (at low and high energies) a Gaussian shape that naturally follows from the Landau model. 
This means that real modeling should be done with advanced three-dimensional hydro codes rather than with simple 
analytic models. The latter can be used to make simple estimates. In particular, the Bjorken model is commonly
used to make an estimate of the initial energy density in the central region at the time when matter becomes
equilibrated. Such estimates usually indicate that this energy is much larger than the critical energy corresponding
to the phase transition. 

%
\subsection{Navier-Stokes hydrodynamics}
\label{sec:1.4}
\sectionmark{Navier-Stokes hydrodynamics}
%
In order to include dissipation in hydrodynamics, one adds a dissipative part
$\Pimunu$ to the perfect-fluid form of the energy-momentum tensor and constructs the complete 
$\Tmunu$ as
\begin{eqnarray}
\Tmunu = \Tmunueq + \Pimunu,
\label{Tmunu}
\end{eqnarray}
where $\Pimunu \unul=0$, which corresponds to the Landau definition of 
the hydrodynamic flow $\umu$
\begin{eqnarray}
T^\mu_{\,\,\,\nu} \unu = \ed\, \umu =  \ed_{\rm eq}\, \umu.
\label{LandFrame}
\end{eqnarray}
It is useful to further decompose $\Pimunu$ into two components,
\bel{pidecomp}
\Pimunu = \pimunu + \Pi \Dmunu,
\eel
where  $\Pi$ is the bulk viscous pressure  (the trace part of $\Pimunu$) and $\pimunu$ 
is the shear stress tensor. The latter is symmetric, $\pimunu=\pinumu$, traceless, 
$\pimumu=0$, and orthogonal to $\umu$, $\pimunu \unul=0$.

In the  Navier-Stokes~\footnote{Claude-Louis Navier, 1785--1836, French engineer and physicist,
Sir George Gabriel Stokes, 1819--1903, Irish physicist and mathematician. The relativistic versions of 
their hydrodynamic equations were introduced by Eckart and Landau.} theory, the  
bulk pressure and shear stress tensor are given by 
the gradients of the flow  vector 
\begin{eqnarray}
\Pi = - \zeta  \p_\mu \umu, \quad  \pimunu =  2 \eta \sigmamunu.
\label{NavierStokes}
\end{eqnarray}
Here $\zeta$ and $\eta$ are the bulk and shear viscosity coefficients,
respectively, and $\sigmamunu$ is the shear flow tensor defined as  
\begin{eqnarray}
\sigmamunu =  \Dmunu_{\alpha \beta} \p^\alpha u^\beta,
\end{eqnarray}
where the projection operator $ \Dmunu_{\alpha \beta} $ has the form
\bel{Deltamnab}
 \Dmunu_{\alpha \beta}  = \f{1}{2} \left(
 \Delta^\mu_{\,\,\,\alpha}  \Delta^\nu_{\,\,\,\beta} +  \Delta^\mu_{\,\,\,\beta}  \Delta^\nu_{\,\,\,\alpha}  \right) 
 - \f{1}{3}  \Dmunu \Delta_{\alpha \beta}.
\eel
The shear viscosity describes the fluid's reaction to the local change of its shape, while the bulk
viscosity describes the reaction to a change of volume. For conformal systems, the bulk
viscosity vanishes
\beq
0 = \Tmumu = \underbrace{ \ed - 3 \peq}_{= 0} - 3 \Pi + \underbrace{\pimumu}_{= 0} = - 3 \Pi, 
\quad \Pi=0.
\eeq
 
In the Navier-Stokes theory the complete energy-momentum tensor has the structure
\begin{eqnarray}
\Tmunu = \Tmunueq + \pimunu + \Pi \Dmunu = \Tmunueq + 2 \eta \sigmamunu -\zeta \theta \Dmunu 
\label{TmunuNS}
\end{eqnarray}
and we use again four equations, 
\bel{conservation1}
\p_\mu \Tmunu = 0,
\eel
for four unknowns: $T(x)$ and three independent components of $\umu(x)$. It turns out, however, that the
use of \rfn{TmunuNS} in the conservation laws \rfn{conservation1} leads to problems connected with
causality and stability of the solutions~\cite{Hiscock:1985zz,Lindblom:1995gp}. Therefore, the framework based on \rfn{TmunuNS}  and 
\rfn{conservation1} should be abandoned in most of practical applications. 

At this place we note that the form \rfn{TmunuNS} can be treated as an expansion of the energy-momentum 
tensor in gradients of $T$ and $u^\mu(x)$ around local equilibrium  up to the terms of the first order in gradients 
\begin{eqnarray}
\Tmunu = \Tmunueq + \pimunu + \Pi \Dmunu = \Tmunueq + \underbrace{2 \eta \sigmamunu -\zeta \theta \Dmunu}_{\hbox{\tiny first order terms in gradients}} .
\label{TmunuNS1}
\end{eqnarray}
One can try to generalise this approach by adding further gradients
\begin{eqnarray}
\Tmunu = \Tmunueq + \underbrace{2 \eta \sigmamunu -\zeta \theta \Dmunu}_{\hbox{\tiny first order terms in gradients}} + \underbrace{.......................}_{\hbox{\tiny second order terms in gradients}} + \ldots
\label{TmunuNS2}
\end{eqnarray}
It turns out, however, that this strategy does not lead to any improvements. As the matter of fact, as
indicated by Heller, Janik, and Witaszczyk~\cite{Heller:2013fn}, this type of the gradient expansion is an asymptotic
series with the convergence radius zero. We shall come back to the formal aspects of the gradient expansion
in Sec.~\ref{sec:4}.

\medskip
Although the approach based on \rfn{TmunuNS}  and \rfn{conservation1} suffers from many
conceptual and practical problems, the form \rfn{TmunuNS} turns out to be a very good approximation 
to many  energy-momentum tensors of  the systems approaching local equilibrium. 
The resolution of this paradox lies in the observation that Eqs.~\rfn{TmunuNS}  and \rfn{conservation1} 
include only hydrodynamic modes, while microscopic theories include both. Therefore,
hydrodynamic models used for interpretation of the data should go beyond the simple
scheme based on \rfn{TmunuNS}  and \rfn{conservation1} .

%
\subsection{Insights from AdS/CFT}
\label{sec:1.5}
\sectionmark{Insights from AdS/CFT}
%

By replacing the quark sector of QCD by a matter sector consisting of 6 scalar fields and 4 Weyl spinor 
fields one obtains a Yang-Mills theory which is conformal and finite ${\cal N}$=\,4 SYM theory.
 In 1990s Maldacena~\cite{Maldacena:1997re} and other authors (Gubser et al.~\cite{Gubser:1998bc}, 
 Witten~\cite{Witten:1998qj})  realized that this quantum field theory, 
taken in the 't Hooft limit, is a string theory that can be studied with the methods of classical
gravity.

Although QCD and  ${\cal N}$=\,4 SYM is quite different (apart from the gluon sector), at sufficiently 
high temperatures these differences become less prominent. In particular, the two theories seem to 
have a small value of the shear viscosity to entropy density ratio. For heavy-ion physics the
important point is that ${\cal N}$=\,4 SYM provides a reliable means of observing how hydrodynamic behavior 
appears in a strongly coupled nonequilibrium system.

In the studies of thermalization, one considers commonly the difference of pressures components for a one-dimensional,
boost-invariant expansion,
\bel{rdef}
R \equiv \f{\pT-\pL}{\peq} .
\eel
Here $\pL$  ($\pT$) is the pressure acting along (transversly to) the beam direction and $\peq$ is the
equilibrium pressure corresponding to the effective temperature $T$ of the system. Introducing a dimensionless
variable
\bel{wdef}
w=\tau T(\tau),
\eel
where $\tau = \sqrt{t^2-z^2}$ is the proper time and $T$ is the effective temperature, and the dimensionless function 
\bel{fowdef1}
f(w) = \f{\tau}{w} \f{dw}{d\tau}  
\eel 
we find~\cite{Heller:2011ju,Heller:2012je}
\bel{fowdef2}
f(w) = \frac{2}{3} + \frac{R}{18}.
\eel 
Computing $f(w)$ within ${\cal N}$=\,4 SYM at late times one finds~\cite{Booth:2009ct}
\bel{fhydro}
f(w) = \f{2}{3}+ \f{1}{9\pi w} +
\f{1-\log 2}{27\pi^2 w^2}+
\f{15-2\pi^2-45\log 2+24 \log^2 2}{972 \pi^3 w^3} +\ldots \ .
\eel
The first term in this expansion corresponds obviously to the perfect-fluid case with $R=0$. The second term
describes a correction due to the shear viscosity that corresponds to the famous Kovtun-Son-Starinets lower bound~\cite{Kovtun:2004de}
\bel{KSS}
\f{\eta}{\sd} = \f{1}{4\pi}.
\eel
We note that this value is smaller than that of any other known substance, including superfluid helium.

The expansion of $f(w)$ contains only inverse integer powers of $w$
\bel{fw}
f(w) = \sum_{n=0}^\infty f_n w^{-n}.
\eel
The computation of further coefficients $f_n$ was done and showed that this series has zero convergence 
radius~\cite{Heller:2013fn}. Thus, the series \rfn{fw} cannot be treated as a good approximation for the exact 
functions $f(w)$. Nevertheless, the first terms become a very good approximation for the exact solutions at late times. This agreement may be attributed to the real hydrodynamic behavior manifested by the system evolving toward local equilibrium.  Interestingly, the agreement between exact solutions and the first terms in \rfn{fw} sets in at rather early times 
(inversely proportional to the initial effective temperature) when the system is anisotropic in the 
momentum space~\cite{Heller:2011ju,Heller:2012je}.  This observation led to the concept of early hydrodynamization 
of matter that replaced the idea of early thermalization. 

Let us note that the idea of early thermalization was introduced in the context of perfect-fluid hydrodynamics used
to describe the RHIC data. The hydrodynamic fits favoured early starting time ($\sim 0.5$ fm/c) of the evolution. Since perfect-fluid hydrodynamics assumes local equilibrium, successful perfect-fluid fits with early starting
time suggested early thermalization.  The situation has changed with the use of  viscous codes.
Although the shear viscosity (divided by the entropy density) is small, the initial flow gradients are large and this leads to large corrections to the
equilibrium pressure. 

\subsection{RTA kinetic equation}
\label{sec:1.6}
\sectionmark{RTA kinetic equation}

Several results discussed below refer to the kinetic-theory approach in the relaxation time approximation. Thus, before
we continue our discussion of heavy-ion phenomenology and various hydrodynamic concepts, it is useful to define this framework.

The Boltzmann kinetic equation in the relaxation time approximation has the form
\bel{rta1}
p^\mu \partial_\mu  f(x,p) =  C[f(x,p)] ,
\eel
where the collision term is given by the expression~\cite{Bhatnagar:1954zz}
\bel{rta2}
C[f] = p^\mu u_\mu \,\, \frac{f^{\rm eq}-f}{\trel}.
\eel
In the case of Boltzmann (classical) statistics, the equilibrium background distribution is 
\bel{feq1}
f^{\rm eq} = \frac{g_s}{(2\pi)^3} \exp\left(-\frac{p^\mu u_\mu}{T}\right) ,
\eel
where $T$ is an effective temperature; $T$ is chosen locally in such a way that $f^{\rm eq}(x,p)$ yields the same 
energy density as $f(x,p)$, which is consistent with \rf{LM0}.

For boost invariant systems, \rf{rta1} is reduced to a simple differential equation~\cite{Baym:1984np,Baym:1985tna,Florkowski:2013lza,Florkowski:2013lya}
\bel{rta3}
\frac{\partial f(\tau, u, v)}{\partial \tau}  = \frac{f^{\rm eq}(\tau, u, v)-f(\tau, u, v)}{\tau_{\rm eq}},
\eel
where we have introduced the variables $u =  t p_L - z E$ and  $v = t E-z p_L$~\cite{Bialas:1984wv,Bialas:1987en}, and the equilibrium distribution function equals
\bel{rta3}
f^{\rm eq}(\tau,u,p_T) = \frac{g_s}{(2\pi)^3} 
\exp\left(- \frac{\sqrt{(u/\tau)^2+p_T^2}}{T}  \right).
\eel
The formal solution of \rf{rta3} is
\bel{rtasol}
f(\tau,u,p_T) &=& D(\tau,\tau_0) f_0(u,p_T) +  \int\limits_{\tau_0}^\tau \frac{d\tau^\prime}{\tau_{\rm eq}(\tau^\prime)} \, D(\tau,\tau^\prime) \, 
f^{\rm eq}(\tau^\prime,u,p_T).
\eel
where $D$ is the damping function
\bel{damp}
D(\tau_2,\tau_1) = \exp\left[-\int\limits_{\tau_1}^{\tau_2}
\frac{d\tau^{\prime\prime}}{\tau_{\rm eq}(\tau^{\prime\prime})} \right]
\eel
and $f_0$ denotes the initial distribution.

\smallskip
The RTA kinetic theory has become a popular tool in many theoretical studies because: 1)~the simple form of the collision term allows for straightforward calculations of the kinetic coefficients~\cite{Anderson:1974a,Anderson:1974b,Czyz:1986mr}, 2)~the knowledge of exact solutions can be used to verify which hydrodynamic model is the best approximation of the kinetic-theory results~\cite{Florkowski:2013lza,Florkowski:2013lya,Denicol:2014xca,Denicol:2014tha}, 3)~the results described in this section can be generalised to the case of finite particle mass~\cite{Florkowski:2014sfa}, which allows for studies of the bulk pressure and shear-bulk coupling effects~\cite{Denicol:2014mca}, 4)~the formal gradient expansion can be done for this model~\cite{Heller:2016rtz} and
compared with similar expansions done for hydrodynamic approaches.

\section{Basic dictionary for phenomenology}
\label{sec:2}
\sectionmark{Basic dictionary for phenomenology}

In this section we very briefly discuss how hydrodynamic modeling of heavy-ion
collision may bring us information about properties of strongly interacting QGP
such as the viscosity coefficients $\eta$ and $\zeta$, and the equation of state. 

\subsection{Glauber model}
\label{sec:2.1}
\sectionmark{Glauber model}

Glauber model treats a nucleus-nucleus collision as a multiple nucleon-nucleon collision process 
(for a review of this approach see, for example, Ref.~\cite{Florkowski:2010zz}). 
In this approach, the nucleon distributions in nuclei are random and given 
by the nuclear density profiles (Woods-Saxon densities for large nuclei),
while the elementary nucleon-nucleon collision is described by the total inelastic cross section $\sigma_{\rm NN}$.
In the original formulation, the Glauber model was applied to elastic collisions only. In this case a nucleon does not change 
its properties in individual collisions, so all nucleon interactions can be well described by the same cross section.
In subsequent applications of the Glauber model to inelastic collisions, one assumes that after a single inelastic collision 
an excited nucleon-like object is created that interacts basically with the same inelastic cross section with other nucleons.
The starting point for the Glauber model is the eikonal approximation -- the classical approximation to the angular momentum
$l$, that is applied to the standard expansion of the elastic scattering amplitude.

The Glauber model can be used to determine the probability of having $n$ inelastic binary nucleon-nucleon collisions in 
a nucleus-nucleus collision at the impact parameter $b$. The Glauber model can be also used to calculate the number of 
nucleons that participate in a collision. To be more precise one distinguishes between the participants which may interact 
elastically and the participants which interact only inelastically. The latter are called the wounded nucleons~\cite{Bialas:1976ed}. 

The results obtained within the Glauber model can be directly used as an input for hydrodynamics. In this case, one
assumes that the initial energy density (or entropy density) is proportional to the density of sources (for particle production)
that are identified with binary collisions and wounded nucleons. In many calculations one simply uses a linear
combinations of the density of binary collisions and wounded nucleons to define the initial entropy density in hydrodynamic
codes.  This procedure assumes implicitly equilibration (in the case where the perfect-fluid hydrodynamics is used) or hydrodynamization (if viscous hydrodynamics is used) of matter at the starting time when 
hydrodynamic equations are initialized. 

\subsection{Harmonic flows}
\label{sec:2.2}
\sectionmark{Harmonic flows}

At high energies one usually distinguishes between the longitudinal direction (along the beam axis)
and the transverse plane (orthogonal to the beam). Then, one introduces the transverse mass
of the produced particles as
\begin{equation}
m_T=\sqrt{m^2+\pt^2} = \sqrt{m^2+p_x^2+p_y^2}.
\label{trmass} 
\end{equation}
The measure of the longitudinal momentum of a particle is rapidty
\begin{equation}
{\tt y}={\frac 12}\ln {\frac{(E+\pl)}{(E-\pl)}}=\hbox{arctanh}
\left( {\f{\pl}E}\right) =\hbox{arctanh}\left( {\tt v}_L \right) .
\label{rap}
\end{equation}
The region ${\tt y} \approx 0$ in the center-of-mass frame of the colliding nuclei is called the 
central region. At ultrarelativistic collisions, in this region the energy density of the produced particles is the highest,
while the baryon number density is the lowest (this feature can be easily explained in the
color-flux-tube models in which the baryon number is carried by the ends of the tubes,
while the baryon free matter is produced by the tube decays, see also our comments about the Bjorken
model in Sec.~\ref{sec:1.3.1}). The central region is
a very suitable place for making basic comparisons between various model predictions
and the data. Since the baryon number density in this region may be neglected, 
the equation of state characterising matter in this region can be obtained directly from
lattice simulations of QCD~\cite{Petreczky:2012rq}.

The produced particles are characterized by their spectra in rapidity and transverse
momentum, which are commonly written in the form
\begin{equation}
\frac{dN}{d{\tt y} d^2 \pt} = \frac{dN}{2 \pi \pt d\pt d{\tt y}}
\left[1 + \sum\limits_{k=1}^\infty 2 v_k \cos\left( k (\phi_p - \Psi_k) \right)\right].
\label{flow-an}
\end{equation}
The angle $\phi_p$ is the azimuthal angle of the three-momentum in the transverse plane, whereas 
the angles $\Psi_k$ define reaction planes (different for each value of $k$). Equation~\rfn{flow-an} is 
nothing other but the Fourier decomposition of the transverse-momentum distribution, with the 
harmonic flow coefficients  $v_k$ characterising the strength of different types of the transverse-momentum 
anisotropy ($v_1$, $v_2$, and $v_3$ are called the directed, elliptic~\cite{Ollitrault:1992bk}, and triangular flow~\cite{Alver:2010gr}, respectively).

A great accomplishment of hydrodynamic modeling of heavy-ion collisions is the explanation of the measured
values of the harmonic flows~\cite{Yan:2017ivm}, in particular of the elliptic flow~\cite{Ollitrault:1992bk}. 
In the hydrodynamic approach, the 
non-zero values of $v_k$ reflect asymmetries in the initial distribution of the energy density --- hydrodynamic
evolution transforms asymmetry of the initial shape of the collision region into asymmetry of the measured
momentum distribution. Very important physical role is played here by fluctuations in the initial state, which lead
to non-zero values of odd harmonic coefficients.

It turns out, that the hydrodynamic predictions of the magnitude of the elliptic flow are sensitive to the
assumed value of the shear viscosity~\cite{Romatschke:2007mq}. This allows for quantitative determination of the shear viscosity
to the entropy density ratio that lies in the range $1/(4\pi) \leq \eta/\sd \leq 2.5/(4\pi)$ \cite{Song:2010mg},
which is very close to the AdS/CFT lower bound~\rfn{KSS}. The measurements of the correlations between produced pions give information about the space-time extensions of the produced system, which are affected by the system's equation of state. This
allows for experimental selection of the proper equation of state, which turns out to be consistent with the lattice simulations of QCD~\cite{Broniowski:2008vp,Pratt:2015zsa}. In the coming years, the bulk viscosity can be also estimated by the experiment data, so that the two main kinetic coefficient of QGP will be determined~\cite{Bass:2017zyn,Alqahtani:2017jwl}.

\section{Viscous fluid dynamics}
\label{sec:3}
\sectionmark{Viscous fluid dynamics}

\subsection{M\"uller-Israel-Stewart theory}
\label{sec:3.1}
\sectionmark{Israel-Stewart theory}

Let us now turn to discussion of different hydrodynamic frameworks.
In the Navier-Stokes theory, the bulk pressure and the shear stress tensor are defined directly by the effective temperature 
and the form of the hydrodynamic flow, see \rf{NavierStokes}.  As we have mentioned above, 
this leads to conceptual and practical problems 
if such a framework is applied to model physical processes (problems with causality and stability). 

In the M\"uller-Israel-Stewart theory~\cite{Muller:1967zza,Israel:1976tn,Israel:1979wp}, the bulk pressure, $\Pi$, 
and  the shear stress tensor, $\pi^{\mu\nu}$, are promoted  to independent dynamic variables which satisfy 
the following two differential equations 
\begin{eqnarray}
\dot{\Pi} + \frac{\Pi}{\tau_{\Pi}} &=&-\beta_{\Pi} \theta,  \label{BULK}\\
\dot{\pi}^{\langle \mu \nu \rangle} + \frac{\pi^{\mu\nu}}{\tau_{\pi}}&=& 
2 \beta_{\pi}  \sigma^{\mu\nu}.  \label{SHEAR}
\end{eqnarray}
Here $\tau_\Pi$ and $\tau_\pi$ are called the relaxation times and the coefficients $\beta_\Pi$ and  $\beta_\pi$
are chosen in such a way that $\eta = \beta_\pi \tau_\pi$ and $\zeta = \beta_\Pi \tau_\Pi$. The dots on the
left-hand sides of \rfn{BULK} and \rfn{SHEAR} denote the convective time derivative $u^\mu \p_\mu$ and
the angular brackets denote contraction with the projector $\Delta^{\mu\nu}_{\alpha\beta}$ defined by \rf{Deltamnab}.
In the case where the terms with the dots are negligible compared to the terms containing the relaxation times,
one reproduces the Navier-Stokes limit.

It is interesting to study initial dynamics of a system that is uniform in space but anisotropic in the momentum
space.  In this case $\theta=0$ and $\sigma^{\mu\nu}=0$ at the initial time, however, initial values of $\Pi$ and 
$\pi^{\mu\nu}$ are, in general, different from zero. According to Eqs.~\rfn{BULK} and \rfn{SHEAR} the initial 
dynamics of such a system is  described by exponential decays of $\Pi$ and $\pi^{\mu\nu}$. The timescales for 
these decays are set by  $\tau_\Pi$ and $\tau_\pi$, respectively. This example illustrates a common situation, 
where initial dynamics of a system is dominated by transient, fast decaying modes. Such modes are known 
in the literature as non-hydrodynamic  modes --- the modes whose frequency $\omega(k)$ does not vanish in 
the limit where the wave vector $k$ vanishes. The modes satisfying the condition $\omega(k) \to 0$ for
$k \to 0$ are known as hydrodynamic modes. 

During the time evolution of a system toward local and possibly global equilibrium, at first the non-hydrodynamic
modes play a role and later the hydrodynamic modes become important. The behavior dominated by
hydrodynamic modes  corresponds to genuine hydrodynamic behavior that can be described by
a few effective parameters such as the viscosity coefficients $\eta$ and $\zeta$. 

Structures such as Eqs.~\rfn{BULK} and \rfn{SHEAR} appear naturally if hydrodynamic equations are derived
from the kinetic theory. As the matter of fact, in such derivations more terms are found than those appearing
in ~\rfn{BULK} and \rfn{SHEAR}. For example, many viscous hydrodynamic models of heavy-ion collisions have the
structure~\cite{Muronga:2001zk,Heinz:2005bw,Bozek:2009dw,Shen:2014vra}
\begin{eqnarray}
\dot{\Pi} +\frac{\Pi}{\tau_{\Pi}} &=&-\beta_{\Pi} \theta 
-\frac{\zeta T}{2 \tau_\Pi }  \Pi
\,\partial_\lambda \left(  \frac{\tau_\Pi}{\zeta T} u^\lambda \right),   \label{BULK1} \\
\dot{\pi}^{\langle\mu\nu\rangle} +\frac{\pi^{\mu\nu}}{\tau_{\pi}} &=& 
2 \beta_{\pi} \sigma^{\mu\nu}
-\frac{\eta T}{2 \tau_\pi} \pi^{\mu\nu}
\,\partial_\lambda \left(  \frac{\tau_\pi}{\eta T} u^\lambda \right).   \label{SHEAR1}
\end{eqnarray}
In the following we shall refer to Eqs.~\rfn{BULK1} and \rfn{SHEAR1} by the acronym MIS.

\subsection{DNMR theory}
\label{sec:3.2}
\sectionmark{DNMR theory}

In order to check which terms should appear in the hydrodynamic equations one should introduce
certain expansion parameters and count their powers. In the context of the kinetic theory (treated
as an underlying microscopic theory for an effective, hydrodynamic description) such an expansion
has been rigorously performed by Denicol, Niemi, Molnar and Rischke (DNMR) 
\cite{Denicol:2010xn,Denicol:2012cn,Denicol:2014loa}.

In their works,  DNMR derive a general expansion of the phase space distribution function 
$\delta f(x,p) = f(x,p)- f_{\rm eq}(x,p)$ in terms of its irreducible moments. In the second step, 
exact equations of motion for these moments are derived.  In general, there is an infinite number of such equations 
and one deals with infinite number of coupled differential equations in order to determine the time evolution of the system. 
However,  substantial reduction of the number of equations is possible,  if the terms are classified 
according to a systematic power-counting scheme in the Knudsen and inverse Reynolds 
numbers~\footnote{The Knudsen number is the ratio of the characteristic microscopic and macroscopic scales
describing the system. The inverse Reynolds numbers count the corrections in  powers of $\sqrt{\pimunu {\pi}_{\mu\nu}}/\peq$ 
and $\Pi/\peq$.}. As long as one keeps terms of second order (in both parameters) the equations of motion can be closed 
and expressed in terms of only 14 dynamical variables.

The DNMR formalism can be applied for a general collision term. In the case where one uses a simplified RTA
form of the collision term, see Sec.~\ref{sec:1.6}, and neglects vorticity, the DNMR equations have the following structure
\begin{eqnarray}
\dot{\Pi} +\frac{\Pi}{\tau_{\Pi}}&=&-\beta_{\Pi}\theta 
-\delta_{\Pi\Pi} \Pi \theta
+ \lambda_{\Pi\pi} \pi^{\mu\nu} \sigma_{\mu \nu},  \label{BULK-DNMR} \\
\dot{\pi}^{\langle\mu\nu\rangle} +\frac{\pi^{\mu\nu}}{\tau_{\pi}}   &=& 
2 \beta_{\pi} \sigma^{\mu\nu}
- \delta_{\pi\pi} \pi^{\mu\nu} \theta 
- \tau_{\pi\pi} \pi_{\gamma}^{\langle\mu} \sigma^{\nu\rangle\gamma} 
+ \lambda_{\pi\Pi} \Pi \sigma^{\mu\nu}.  \label{SHEAR-DNMR}
\end{eqnarray}
The terms $\delta_{\Pi\Pi}$, $\lambda_{\Pi\pi}$, $\delta_{\pi\pi}$, $\tau_{\pi\pi}$, $\lambda_{\pi\Pi}$ 
are new kinetic coefficients (compared to MIS). Their form follows directly from the RTA collision term. 
An interesting feature of Eqs.~\rfn{BULK-DNMR} and \rfn{SHEAR-DNMR} is the presence of coupling between 
the bulk and shear sectors --- terms proportional to $\lambda_{\Pi\pi}$ and $\lambda_{\pi\Pi}$ 
\cite{Denicol:2014mca}.

\subsection{BRSSS theory}
\label{sec:3.3}
\sectionmark{BRSSS theory}

The DNMR approach constructs hydrodynamic equations in a direct relation to the kinetic theory. It is possible,
however, to construct hydrodynamic equations without reference to any microscopic model or theory. An example of
such a formulation for conformal systems is the method worked out by Baier, Romatschke, Son, Starinets, and 
Stephanov (BRSSS)~\cite{Baier:2007ix}. In this formulation one constructs first the shear stress tensor out of  gradients of the 
effective temperature $T$ and the flow vector $u^\mu$. This construction is based on the gradient expansion 
and includes all terms up to the second order which are allowed by the Lorentz and conformal symmetry.  
Among several allowed terms, the expression for $\pi^{\mu\nu}$ contains the term $\dot{\sigma}^{\mu\nu}$.
Using the first order relation between $\pi^{\mu\nu}$ and $\sigma^{\mu\nu}$, namely the Navier-Stokes
relation $\pi^{\mu\nu} = 2 \eta \sigma^{\mu\nu}$, one can replace the term $\dot{\sigma}^{\mu\nu}$
by an expression containing $\dot{\pi}^{\mu\nu}$. In this way, one obtains a dynamic equation for 
the shear stress tensor, which has the form familiar from the MIS or DNMR theories. 

What is important in the BRSSS approach is that it does not refer directly to any microscopic theory.
The kinetic coefficients that appear in this approach should be matched to any underlying theory or the
experiment.  If one neglects the terms icluding vorticity and space-time curvature, the BRSSS equation 
take the form~\cite{Baier:2007ix}
\begin{eqnarray}
\Pi&=& 0,  \label{BULK-BRSSS}\\
\dot{\pi}^{\langle\mu\nu\rangle} +\frac{\pi^{\mu\nu}}{\tau_{\pi}}&=& 
2\textcolor{black}{\beta_{\pi}}\sigma^{\mu\nu}
-\frac{4}{3} \pi^{\mu\nu} \theta  
+\frac{\lambda_1}{\tau_\pi \eta^2} \pi^{\langle \mu}_{\,\,\lambda} \pi^{\nu \rangle \lambda}.
 \label{SHEAR-BRSSS}
\end{eqnarray}
Equation \rfn{BULK-BRSSS} reflects the fact that we deal with conformal systems. The term $\lambda_1$ is a new kinetic coefficient.

\subsection{Anisotropic hydrodynamics}
\label{sec:3.4}
\sectionmark{Anisotropic hydrodynamics}

As we have discussed above,  viscous hydrodynamics can be derived as expansion of the underlying microscopic kinetic theory in the  Knudsen and  inverse Reynolds numbers around the local equilibrium state~\cite{Denicol:2014loa}. This type of expansion may be questioned in the situation where space-time gradients and/or deviations from the local equilibrium are large.  The goal of the anisotropic hydrodynamics program is to create a dissipative hydrodynamics framework that is better suited to deal with such cases (for a recent review see~\cite{Alqahtani:2017mhy}).

The initial ideas of anisotropic hydrodynamics were introduced in Refs.~\cite{Florkowski:2010cf,Martinez:2010sc}, see also \cite{Barz:1987pq,Kampfer:1990qg}. They were restricted to the boost-invariant, one-dimensional case. The framework of \cite{Florkowski:2010cf} was based on the energy-momentum conservation law and used a specific ansatz for the entropy source term that defined the off-equilibrium dynamics. The approach of \cite{Martinez:2010sc} was based on the kinetic theory, and employed the zeroth and first moments of the Boltzmann kinetic equation in the relaxation time approximation.

Subsequent developments of anisotropic hydrodynamics were based mainly on the kinetic theory and can be classified as perturbative and non-perturbative schemes. In the perturbative approach~\cite{Bazow:2013ifa,Molnar:2016vvu,Molnar:2016gwq} one assumes that the phase-space distribution function has the form  $f = f_{RS} + \delta f $, where $f_{RS}$ is the leading order described by the Romatschke-Strickland form~\cite{Romatschke:2003ms}, accounting for the difference between  the longitudinal and transverse pressures, while $\delta f$  represents a correction.  In this case, advanced methods of traditional viscous hydrodynamics (following DNMR) are used to restrict the form of $\delta f$ and to derive hydrodynamic equations. In this way non-trivial dynamics may be included in the transverse plane and, more generally, in the full (3+1)D case.
In the non-perturbative approach one starts with the decomposition  $f = f_{\rm aniso} +  \delta f $, where $f_{\rm aniso}$ is the leading order distribution function given by the generalised RS form. In this case, all effects due to anisotropy are included in the leading order, while the correction term $ \delta f$ is typically neglected. The latest development in this direction is known
as the anisotropic matching principle~\cite{Tinti:2015xwa}.

Anisotropic and viscous hydrodynamics predictions have been checked against exact solutions available for the Boltzmann kinetic equation in the relaxation time approximation. Such studies have been done for the one-dimensional Bjorken geometry~\cite{Florkowski:2013lza,Florkowski:2013lya,Florkowski:2014sfa} and for the Gubser flow that includes transverse expansion~\cite{Denicol:2014xca,Denicol:2014tha}. Such studies showed that anisotropic hydrodynamics better reproduces the results of the underlying kinetic theory than the standard viscous hydrodynamics. In addition, important constraints on the structure of the hydrodynamic equations and the form of the kinetic coefficients have been obtained within such studies. 

\section{Gradient expansion}
\label{sec:4}
\sectionmark{Gradient expansion}

\subsection{Formal aspects}
\label{sec:4.1}
\sectionmark{Formal aspects}

\begin{figure}[t] 
\begin{center}
\includegraphics[angle=0,width=0.7\textwidth]{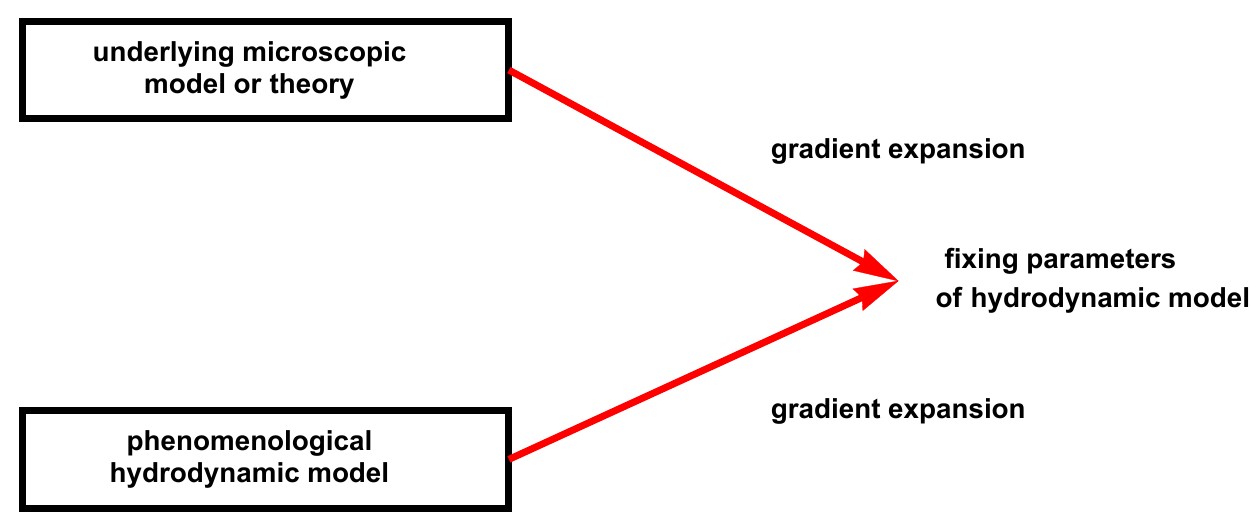}
\end{center} 
 \caption{Schematic illustration of the gradient expansion as a formal tool that can be used for comparisons
 between microscopic theories and phenomenological hydrodynamic models. Such comparisons can be 
 used to transfer the knowledge about the kinetic coefficients from the microscopic theory to hydrodynamic
 frameworks. }
 \label{fig:ge2}
\end{figure}

\begin{figure}[t] 
\begin{center}
\includegraphics[angle=0,width=0.8\textwidth]{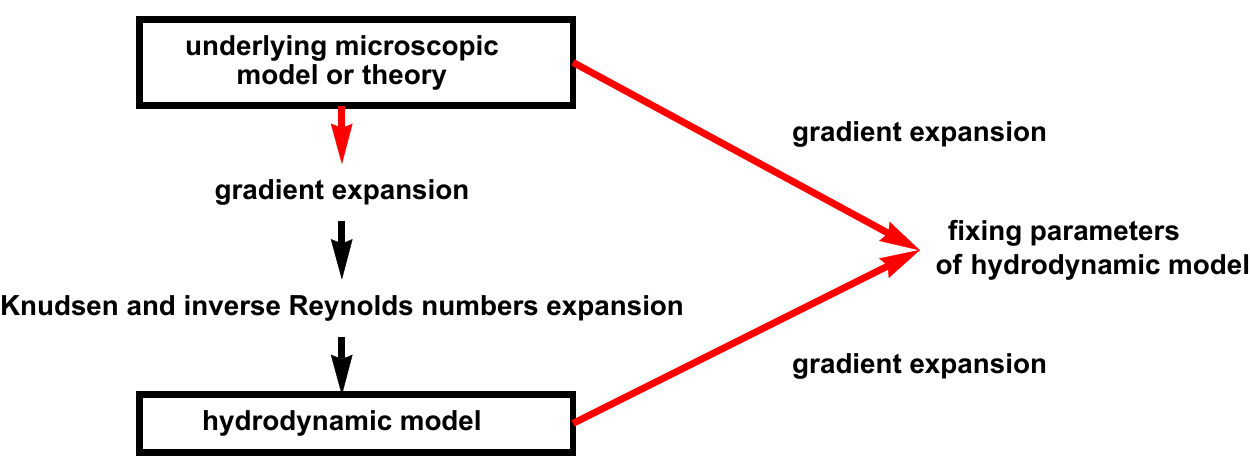}
\end{center} 
 \caption{Schematic view how the gradient expansion can be used to
 check the internal consistency between kinetic theory and the hydrodynamic
 model that has been derived as its approximation (for example by making
 a simultaneous expansion in the Knudsen and inverse Reynolds numbers.)}
 \label{fig:ge1}
\end{figure}

It is very much convenient to have a formal method that allows for making comparisons between
different hydrodynamical models and, more generally, between hydrodynamic models
and close-to-equilibrium behavior of microscopic theories for which hydrodynamic
models are regarded as good approximations.  Such method exists and is based on the
formal expansion of the energy-momentum tensor in gradients of $T$ and $u^\mu$ (around
its local equilibrium form),
\begin{eqnarray}
T^{\mu\nu} &=& T^{\mu\nu}_{\rm eq} + \hbox{powers of gradients of } T \hbox{ and } u^\mu .
\label{gradexp}
\end{eqnarray}
It is important to emphasise that elements of the gradient expansion are present
in various methods used to derive hydrodynamic equations from the underlying
microscopic theories. For example, the counting in the Knudsen number or the
first step in the BRSSS method refer to the number of gradients of $T$ and $u^\mu$. 
However, such derivations include usually other arguments or assumptions
that allow for construction of a self-consistent system of equations. In contrast
to such derivations, the gradient expansion of the energy-momentum tensor \rfn{gradexp}
is not useful for finding approximate solutions of the microscopic theory.
It should be regarded rather as a formal tool to make comparisons between
different theories and to check their close-to-equilibrium behavior.
This is illustrated schematically in Figs. \ref{fig:ge2} and \ref{fig:ge1}.

In the first case, see Fig. \ref{fig:ge2}, one performs the gradient expansion of the energy-momentum
tensor for a specified microscopic theory and a phenomenological
hydrodynamic model (the latter can be taken, for example, in the BRSSS formulation
that has unspecified values of the kinetic coefficients). 
The comparison of the two expansions allows for fixing the values of kinetic
coefficients in the hydrodynamic model that can be later used to
approximate the close-to-equilibrium behavior of the microscopic theory.
This strategy has been used in heavy-ion modeling by using the value
of the shear viscosity obtained from the SYM theory in the hydrodynamic codes.
In the second case, see Fig. \ref{fig:ge1}, one can use the gradient expansion 
to check the internal consistsncy between the underlying theory and the
hydrodynamic model which has been constructed as its close-to-equilibrium
approximation.

\subsection{RTA kinetic model with Bjorken geometry}
\label{sec:4.2}
\sectionmark{Bjorken geometry}

Performing the gradient expansion in a general case is quite complicated but the calculations 
become doable in the cases where the flow pattern is restricted by certain symmetries.
This is the case of the Bjorken-type geometry, where expansion of matter
is one-dimensional and boost invariant.  

In this section we consider the case where the underlying theory corresponds to the conformal
RTA kinetic model. For this model the gradient expansion has been performed recently
by Heller, Kurkela and Spalinski~\cite{Heller:2016rtz}. Their results can be compared with the expansions
done for the hydrodynamic models, for which we use the MIS, DNMR and BRSSS
versions. For the Bjorken geometry the only independent component of the shear stress tensor 
is its longitudinal component \mbox{$\phi = -\pi^{zz}$}. The equation expressing the energy and momentum 
conservation is common to all the hydrodynamic frameworks and reads
\beal{miseqn}
\tau  \dot{\ed} &=& - \frac{4}{3}\ed + \phi.
\eea
On the other hand, the dynamic equation for $\phi$ depends on the hydrodynamic framework. For the
MIS case with the RTA kinetic equation one finds
\beal{miseqn}
\tau_\pi \dot{\phi} &=& 
\frac{4 \eta}{3 \tau } 
- \frac{4 \tau_\pi\phi}{3 \tau }
- \phi.
\eea
For the DNMR framework, again with the RTA kinetic equation, we have
\beal{dnmreqn}
\tau_\pi \dot{\phi} &=& 
\frac{4 \eta}{3 \tau } 
- \frac{38}{21} \f{\tau_\pi\phi}{\tau }
- \phi.
\eea
Finally, for the BRSSS approach one finds
\beal{brssseqn}
\tau_\pi \dot{\phi} &=& 
\frac{4 \eta}{3 \tau } 
- \frac{\lambda_1\phi^2}{2 \eta^2}
- \frac{4 \tau_\pi\phi}{3 \tau }
- \phi.
\eea

For conformal systems, the relaxation time should scale inversely with the temperature, namely
\bel{contrel1}
\trel = \f{c}{T},
\eel
where $c$ is a constant. For the RTA kinetic equation one can connect the relaxation time with the ratio
of the shear viscosity to the entropy density~\cite{Anderson:1974a}
\bel{contrel2}
c = \f{5 \eta}{\sd},
\eel
thus, adopting the value of the $\eta/\sd$ ratio (for example as $1/(4\pi)$, see \rf{KSS})  we fix $c$. 

Constructing the gradient expansion for hydrodynamic models we expand
$T$, $\ed$ and $\phi$ around the Bjorken solution. For example, in the case of $T$ we use the series
\begin{eqnarray}
T &=& T_0 \left(\frac{\tau_0}{\tau}\right)^{1/3} \left( 1 + \sum_{n=1}^\infty
\left(\frac{c}{T_0\tau_0}\right)^n t_n \left(\frac{\tau_0}{\tau}\right)^{2 n/3} \right),
\end{eqnarray}
where $\tau$ is the proper time and $\tau_0$ is the initial proper time for which the initial Bjorken temperature is $T_0$.  Using similar expansions for $\ed$ and $\phi$ we find the expansion coefficients and may construct the gradient expansion of $T^{\mu\nu}$.
However, as demonstrated in~\cite{Heller:2011ju,Heller:2012je} it is better to analyze the expansion of the function $f(w)$ defined in Sec.~\ref{sec:1.5}. 

The results for the coefficients $f_n$ obtained for the RTA kinetic model and various hydrodynamic formulations are shown in Table~1. We can see that both BRSSS and DNMR agree with the exact results up to the second order in $n$~\footnote{We note that it is possible to get the agreement up to the third order if the third-order dissipative hydrodynamics formulated in~\cite{Jaiswal:2013vta} is used.}. There are, however, differences in obtaining such agreement in the two cases. For the BRSSS framework the agreement has been achieved by using the same value of the viscosity (as in the RTA model) and by fitting the $\lambda_1$ parameter in \rfn{brssseqn} --- we remember that BRSSS does not specify the kinetic coefficients, thus, they can be adjusted to any microscopic model. For the DNMR approach the agreement is a consquence that the kinetic coefficients used in \rfn{dnmreqn} follow from the RTA model. Table 1 shows also the results for MIS, which agree only up to the first order. This indicates that the MIS framework is incomplete (in the second order).

\begin{table}[t]
  \label{tab:table2}
  \begin{center}
  \begin{tabular}{|c|c|c|c|c|}
    \hline
    $n$ & RTA & BRSSS & DNMR & MIS \\
    \hline
    $0$ & $2/3$ & $2/3$ & $2/3$ & $2/3$ \\
    \hline
    $1$ &  $4/45$ &  $4/45$ & $4/45$ & $4/45$ \\
    \hline    
    $2$ & $16/945 $ & $16/945$ & $16/945$ & $8/135 $\\
    \hline
    $3$ & $- 208/4725$ & $-1712/99225$  & -$304/33075$ & $112/2025$\\
        \hline
    $3$ & $-0.044$ & $-0.017$  & -$0.009$ & $0.055$\\
    \hline
  \end{tabular}
  \end{center}
\caption{Expansion coefficients $f_n$ for the RTA kinetic model and various hydrodynamic frameworks.  }
\end{table}

\section{Closing remarks}
\label{sec:5}
\sectionmark{Closing remarks}

Success of relativistic viscous hydrodynamics as a basic theoretical tool used to model heavy-ion collisions
triggered broad interest in formal aspects of this framework and relations between hydrodynamic models and
underlying microscopic theories. In these lectures we discussed a few topics related to formal gradient
expansion and mutual relations between different hydrodynamic models. For a discussion of other issues
we refer to the articles mentioned in Introduction.

\begin{acknowledgement}
I thank Chihiro Sasaki, David Blaschke, Krzysztof Redlich and Ludwik Turko, the organizers of the \emph{53rd Karpacz Winter School of Theoretical Physics}, for very kind hospitality. I also thank Micha\l{}~P.~Heller and Micha\l{} Spali\'nski for illuminating discussions on several topics presented in this paper.  This work was supported in part by the by the Polish National Science Center Grant No. 2016/23/B/ST2/00717 and by the ExtreMe Matter Institute EMMI at the GSI Helmholtzzentrum f\"ur Schwerionenforschung, Darmstadt, Germany.
\end{acknowledgement}

  \bibliographystyle{utphys}
 \bibliography{hydroreview}

\providecommand{\href}[2]{#2}\begingroup\raggedright\begin{thebibliography}{10}

\bibitem{Florkowski:2017olj}
W.~Florkowski, M.~P. Heller, and M.~Spalinski, ``{New theories of relativistic
  hydrodynamics in the LHC era},''
\href{http://arxiv.org/abs/1707.02282}{{\tt arXiv:1707.02282 [hep-ph]}}.

\bibitem{Romatschke:2009im}
P.~Romatschke, ``{New Developments in Relativistic Viscous Hydrodynamics},''
  \href{http://dx.doi.org/10.1142/S0218301310014613}{{\em Int. J. Mod. Phys.}
  {\bf E19} (2010)  1--53},
\href{http://arxiv.org/abs/0902.3663}{{\tt arXiv:0902.3663 [hep-ph]}}.

\bibitem{Jeon:2015dfa}
S.~Jeon and U.~Heinz, ``{Introduction to Hydrodynamics},''
  \href{http://dx.doi.org/10.1142/S0218301315300106}{{\em Int. J. Mod. Phys.}
  {\bf E24} (2015) no.~10, 1530010},
\href{http://arxiv.org/abs/1503.03931}{{\tt arXiv:1503.03931 [hep-ph]}}.

\bibitem{Jaiswal:2016hex}
A.~Jaiswal and V.~Roy, ``{Relativistic hydrodynamics in heavy-ion collisions:
  general aspects and recent developments},''
  \href{http://dx.doi.org/10.1155/2016/9623034}{{\em Adv. High Energy Phys.}
  {\bf 2016} (2016)  9623034},
\href{http://arxiv.org/abs/1605.08694}{{\tt arXiv:1605.08694 [nucl-th]}}.

\bibitem{Alqahtani:2017mhy}
M.~Alqahtani, M.~Nopoush, and M.~Strickland, ``{Relativistic anisotropic
  hydrodynamics},''
\href{http://arxiv.org/abs/1712.03282}{{\tt arXiv:1712.03282 [nucl-th]}}.

\bibitem{Yan:2017ivm}
L.~Yan, ``{A flow paradigm in heavy-ion collisions},''
\href{http://arxiv.org/abs/1712.04580}{{\tt arXiv:1712.04580 [nucl-th]}}.

\bibitem{LLfluid}
L.~D. Landau and E.~M. Lifshitz, {\em Fluid Mechanics, Second Edition: Volume 6
  (Course of Theoretical Physics)}.
\newblock 1987.

\bibitem{rezzolla2013relativistic}
L.~Rezzolla and O.~Zanotti, {\em Relativistic Hydrodynamics}.
\newblock OUP Oxford, 2013.

\bibitem{Yagi:2005yb}
K.~Yagi, T.~Hatsuda, and Y.~Miake, ``{Quark-gluon plasma: From big bang to
  little bang},''
{\em Camb. Monogr. Part. Phys. Nucl. Phys. Cosmol.} {\bf 23} (2005)  1--446.

\bibitem{Vogt:2007zz}
R.~Vogt, {\em {Ultrarelativistic heavy-ion collisions}}.
\newblock Elsevier, Amsterdam,
2007.
\newblock

\bibitem{Florkowski:2010zz}
W.~Florkowski, {\em {Phenomenology of Ultra-Relativistic Heavy-Ion
  Collisions}}.
\newblock World Scientific, Singapore,
2010.
\newblock

\bibitem{CasalderreySolana:2011us}
J.~Casalderrey-Solana, H.~Liu, D.~Mateos, K.~Rajagopal, and U.~A. Wiedemann,
  {\em {Gauge/String Duality, Hot QCD and Heavy Ion Collisions}}.
\newblock 2011.
\newblock
\href{http://arxiv.org/abs/1101.0618}{{\tt arXiv:1101.0618 [hep-th]}}.
\newblock

\bibitem{DeWolfe:2013cua}
O.~DeWolfe, S.~S. Gubser, C.~Rosen, and D.~Teaney, ``{Heavy ions and string
  theory},'' \href{http://dx.doi.org/10.1016/j.ppnp.2013.11.001}{{\em Prog.
  Part. Nucl. Phys.} {\bf 75} (2014)  86--132},
\href{http://arxiv.org/abs/1304.7794}{{\tt arXiv:1304.7794 [hep-th]}}.

\bibitem{Heller:2016gbp}
M.~P. Heller, ``{Holography, Hydrodynamization and Heavy-Ion Collisions},''
  \href{http://dx.doi.org/10.5506/APhysPolB.47.2581}{{\em Acta Phys. Polon.}
  {\bf B47} (2016)  2581},
\href{http://arxiv.org/abs/1610.02023}{{\tt arXiv:1610.02023 [hep-th]}}.

\bibitem{Petreczky:2012rq}
P.~Petreczky, ``{Lattice QCD at non-zero temperature},''
  \href{http://dx.doi.org/10.1088/0954-3899/39/9/093002}{{\em J. Phys.} {\bf
  G39} (2012)  093002},
\href{http://arxiv.org/abs/1203.5320}{{\tt arXiv:1203.5320 [hep-lat]}}.

\bibitem{Floris:2014pta}
M.~Floris, ``{Hadron yields and the phase diagram of strongly interacting
  matter},'' \href{http://dx.doi.org/10.1016/j.nuclphysa.2014.09.002}{{\em
  Nucl. Phys.} {\bf A931} (2014)  103--112},
\href{http://arxiv.org/abs/1408.6403}{{\tt arXiv:1408.6403 [nucl-ex]}}.

\bibitem{Romatschke:2016hle}
P.~Romatschke, ``{Do nuclear collisions create a locally equilibrated
  quark-gluon plasma?},''
  \href{http://dx.doi.org/10.1140/epjc/s10052-016-4567-x}{{\em Eur. Phys. J.}
  {\bf C77} (2017) no.~1, 21},
\href{http://arxiv.org/abs/1609.02820}{{\tt arXiv:1609.02820 [nucl-th]}}.

\bibitem{Spalinski:2016fnj}
M.~Spalinski, ``{Small systems and regulator dependence in relativistic
  hydrodynamics},'' \href{http://dx.doi.org/10.1103/PhysRevD.94.085002}{{\em
  Phys. Rev.} {\bf D94} (2016) no.~8, 085002},
\href{http://arxiv.org/abs/1607.06381}{{\tt arXiv:1607.06381 [nucl-th]}}.

\bibitem{Romatschke:2017vte}
P.~Romatschke, ``{Far From Equilibrium Fluid Dynamics},''
\href{http://arxiv.org/abs/1704.08699}{{\tt arXiv:1704.08699 [hep-th]}}.

\bibitem{Landau:1953gs}
L.~D. Landau, ``On the multiparticle production in high-energy collisions,''
{\em Izv. Akad. Nauk SSSR Ser. Fiz.} {\bf 17} (1953)  51--64.

\bibitem{Bjorken:1982qr}
J.~D. Bjorken, ``{Highly Relativistic Nucleus-Nucleus Collisions: The Central
  Rapidity Region},''
\href{http://dx.doi.org/10.1103/PhysRevD.27.140}{{\em Phys. Rev.} {\bf D27}
  (1983)  140--151}.

\bibitem{Hiscock:1985zz}
W.~A. Hiscock and L.~Lindblom, ``{Generic instabilities in first-order
  dissipative relativistic fluid theories},''
\href{http://dx.doi.org/10.1103/PhysRevD.31.725}{{\em Phys.Rev.} {\bf D31}
  (1985)  725--733}.

\bibitem{Lindblom:1995gp}
L.~Lindblom, ``{The Relaxation effect in dissipative relativistic fluid
  theories},'' \href{http://dx.doi.org/10.1006/aphy.1996.0036}{{\em Annals
  Phys.} {\bf 247} (1996)  1},
\href{http://arxiv.org/abs/gr-qc/9508058}{{\tt arXiv:gr-qc/9508058 [gr-qc]}}.

\bibitem{Heller:2013fn}
M.~P. Heller, R.~A. Janik, and P.~Witaszczyk, ``{Hydrodynamic Gradient
  Expansion in Gauge Theory Plasmas},''
  \href{http://dx.doi.org/10.1103/PhysRevLett.110.211602}{{\em Phys.Rev.Lett.}
  {\bf 110} (2013) no.~21, 211602},
\href{http://arxiv.org/abs/1302.0697}{{\tt arXiv:1302.0697 [hep-th]}}.

\bibitem{Maldacena:1997re}
J.~M. Maldacena, ``{The Large N limit of superconformal field theories and
  supergravity},'' {\em Adv.Theor.Math.Phys.} {\bf 2} (1998)  231--252,
\href{http://arxiv.org/abs/hep-th/9711200}{{\tt arXiv:hep-th/9711200
  [hep-th]}}.

\bibitem{Gubser:1998bc}
S.~Gubser, I.~R. Klebanov, and A.~M. Polyakov, ``{Gauge theory correlators from
  noncritical string theory},''
  \href{http://dx.doi.org/10.1016/S0370-2693(98)00377-3}{{\em Phys.Lett.} {\bf
  B428} (1998)  105--114},
\href{http://arxiv.org/abs/hep-th/9802109}{{\tt arXiv:hep-th/9802109
  [hep-th]}}.

\bibitem{Witten:1998qj}
E.~Witten, ``{Anti-de Sitter space and holography},'' {\em
  Adv.Theor.Math.Phys.} {\bf 2} (1998)  253--291,
\href{http://arxiv.org/abs/hep-th/9802150}{{\tt arXiv:hep-th/9802150
  [hep-th]}}.

\bibitem{Heller:2011ju}
M.~P. Heller, R.~A. Janik, and P.~Witaszczyk, ``{The characteristics of
  thermalization of boost-invariant plasma from holography},''
  \href{http://dx.doi.org/10.1103/PhysRevLett.108.201602}{{\em Phys.Rev.Lett.}
  {\bf 108} (2012)  201602},
\href{http://arxiv.org/abs/1103.3452}{{\tt arXiv:1103.3452 [hep-th]}}.

\bibitem{Heller:2012je}
M.~P. Heller, R.~A. Janik, and P.~Witaszczyk, ``{A numerical relativity
  approach to the initial value problem in asymptotically Anti-de Sitter
  spacetime for plasma thermalization - an ADM formulation},''
  \href{http://dx.doi.org/10.1103/PhysRevD.85.126002}{{\em Phys.Rev.} {\bf D85}
  (2012)  126002},
\href{http://arxiv.org/abs/1203.0755}{{\tt arXiv:1203.0755 [hep-th]}}.

\bibitem{Booth:2009ct}
I.~Booth, M.~P. Heller, and M.~Spalinski, ``{Black brane entropy and
  hydrodynamics: The Boost-invariant case},''
  \href{http://dx.doi.org/10.1103/PhysRevD.80.126013}{{\em Phys. Rev.} {\bf
  D80} (2009)  126013},
\href{http://arxiv.org/abs/0910.0748}{{\tt arXiv:0910.0748 [hep-th]}}.

\bibitem{Kovtun:2004de}
P.~Kovtun, D.~T. Son, and A.~O. Starinets, ``{Viscosity in strongly interacting
  quantum field theories from black hole physics},''
  \href{http://dx.doi.org/10.1103/PhysRevLett.94.111601}{{\em Phys. Rev. Lett.}
  {\bf 94} (2005)  111601},
\href{http://arxiv.org/abs/hep-th/0405231}{{\tt arXiv:hep-th/0405231
  [hep-th]}}.

\bibitem{Bhatnagar:1954zz}
P.~L. Bhatnagar, E.~P. Gross, and M.~Krook, ``{A Model for Collision Processes
  in Gases. 1. Small Amplitude Processes in Charged and Neutral One-Component
  Systems},''
\href{http://dx.doi.org/10.1103/PhysRev.94.511}{{\em Phys. Rev.} {\bf 94}
  (1954)  511--525}.

\bibitem{Baym:1984np}
G.~Baym, ``{Thermal equilibration in ultra relativistic heavy-ion
  collisions},''
\href{http://dx.doi.org/10.1016/0370-2693(84)91863-X}{{\em Phys. Lett.} {\bf
  B138} (1984)  18--22}.

\bibitem{Baym:1985tna}
G.~Baym, ``{Entropy production and the evolution of ultra relativistic
  heavy-ion collisions },''
\href{http://dx.doi.org/10.1016/0375-9474(84)90573-6}{{\em Nucl. Phys.} {\bf
  A418} (1984)  525C--537C}.

\bibitem{Florkowski:2013lza}
W.~Florkowski, R.~Ryblewski, and M.~Strickland, ``{Anisotropic Hydrodynamics
  for Rapidly Expanding Systems},''
  \href{http://dx.doi.org/10.1016/j.nuclphysa.2013.08.004}{{\em Nucl. Phys.}
  {\bf A916} (2013)  249--259},
\href{http://arxiv.org/abs/1304.0665}{{\tt arXiv:1304.0665 [nucl-th]}}.

\bibitem{Florkowski:2013lya}
W.~Florkowski, R.~Ryblewski, and M.~Strickland, ``{Testing viscous and
  anisotropic hydrodynamics in an exactly solvable case},''
  \href{http://dx.doi.org/10.1103/PhysRevC.88.024903}{{\em Phys. Rev.} {\bf
  C88} (2013)  024903},
\href{http://arxiv.org/abs/1305.7234}{{\tt arXiv:1305.7234 [nucl-th]}}.

\bibitem{Bialas:1984wv}
A.~Bialas and W.~Czyz, ``{Boost Invariant Boltzmann-vlasov Equations for
  Relativistic Quark - Anti-quark Plasma},''
\href{http://dx.doi.org/10.1103/PhysRevD.30.2371}{{\em Phys. Rev.} {\bf D30}
  (1984)  2371}.

\bibitem{Bialas:1987en}
A.~Bialas, W.~Czyz, A.~Dyrek, and W.~Florkowski, ``{Oscillations of Quark -
  Gluon Plasma Generated in Strong Color Fields},''
\href{http://dx.doi.org/10.1016/0550-3213(88)90035-1}{{\em Nucl. Phys.} {\bf
  B296} (1988)  611--624}.

\bibitem{Anderson:1974a}
J.~L. Anderson and H.~R. Witting, ``{A relativistic relaxation-time model for
  the Boltzmann equation},'' {\em Physica} {\bf 74} (1974)  466.

\bibitem{Anderson:1974b}
J.~L. Anderson and H.~R. Witting, ``{Relativistic quantum transport
  coefficients},'' {\em Physica} {\bf 74} (1974)  489.

\bibitem{Czyz:1986mr}
W.~Czyz and W.~Florkowski, ``{Kinetic Coefficients for Quark - Anti-quark
  Plasma},''
{\em Acta Phys. Polon.} {\bf B17} (1986)  819--837.

\bibitem{Denicol:2014xca}
G.~S. Denicol, U.~W. Heinz, M.~Martinez, J.~Noronha, and M.~Strickland, ``{New
  Exact Solution of the Relativistic Boltzmann Equation and its Hydrodynamic
  Limit},'' \href{http://dx.doi.org/10.1103/PhysRevLett.113.202301}{{\em Phys.
  Rev. Lett.} {\bf 113} (2014) no.~20, 202301},
\href{http://arxiv.org/abs/1408.5646}{{\tt arXiv:1408.5646 [hep-ph]}}.

\bibitem{Denicol:2014tha}
G.~S. Denicol, U.~W. Heinz, M.~Martinez, J.~Noronha, and M.~Strickland,
  ``{Studying the validity of relativistic hydrodynamics with a new exact
  solution of the Boltzmann equation},''
  \href{http://dx.doi.org/10.1103/PhysRevD.90.125026}{{\em Phys. Rev.} {\bf
  D90} (2014) no.~12, 125026},
\href{http://arxiv.org/abs/1408.7048}{{\tt arXiv:1408.7048 [hep-ph]}}.

\bibitem{Florkowski:2014sfa}
W.~Florkowski, E.~Maksymiuk, R.~Ryblewski, and M.~Strickland, ``{Exact solution
  of the (0+1)-dimensional Boltzmann equation for a massive gas},''
  \href{http://dx.doi.org/10.1103/PhysRevC.89.054908}{{\em Phys. Rev.} {\bf
  C89} (2014) no.~5, 054908},
\href{http://arxiv.org/abs/1402.7348}{{\tt arXiv:1402.7348 [hep-ph]}}.

\bibitem{Denicol:2014mca}
G.~S. Denicol, W.~Florkowski, R.~Ryblewski, and M.~Strickland, ``{Shear-bulk
  coupling in nonconformal hydrodynamics},''
  \href{http://dx.doi.org/10.1103/PhysRevC.90.044905}{{\em Phys. Rev.} {\bf
  C90} (2014) no.~4, 044905},
\href{http://arxiv.org/abs/1407.4767}{{\tt arXiv:1407.4767 [hep-ph]}}.

\bibitem{Heller:2016rtz}
M.~P. Heller, A.~Kurkela, and M.~Spalinski, ``{Hydrodynamization and transient
  modes of expanding plasma in kinetic theory},''
\href{http://arxiv.org/abs/1609.04803}{{\tt arXiv:1609.04803 [nucl-th]}}.

\bibitem{Bialas:1976ed}
A.~Bialas, M.~Bleszynski, and W.~Czyz, ``{Multiplicity Distributions in
  Nucleus-Nucleus Collisions at High-Energies},''
\href{http://dx.doi.org/10.1016/0550-3213(76)90329-1}{{\em Nucl. Phys.} {\bf
  B111} (1976)  461--476}.

\bibitem{Ollitrault:1992bk}
J.-Y. Ollitrault, ``{Anisotropy as a signature of transverse collective
  flow},''
\href{http://dx.doi.org/10.1103/PhysRevD.46.229}{{\em Phys. Rev.} {\bf D46}
  (1992)  229--245}.

\bibitem{Alver:2010gr}
B.~Alver and G.~Roland, ``{Collision geometry fluctuations and triangular flow
  in heavy-ion collisions},''
  \href{http://dx.doi.org/10.1103/PhysRevC.82.039903,
  10.1103/PhysRevC.81.054905}{{\em Phys. Rev.} {\bf C81} (2010)  054905},
  \href{http://arxiv.org/abs/1003.0194}{{\tt arXiv:1003.0194 [nucl-th]}}.
[Erratum: Phys. Rev.C82,039903(2010)].

\bibitem{Romatschke:2007mq}
P.~Romatschke and U.~Romatschke, ``{Viscosity Information from Relativistic
  Nuclear Collisions: How Perfect is the Fluid Observed at RHIC?},''
  \href{http://dx.doi.org/10.1103/PhysRevLett.99.172301}{{\em Phys. Rev. Lett.}
  {\bf 99} (2007)  172301},
\href{http://arxiv.org/abs/0706.1522}{{\tt arXiv:0706.1522 [nucl-th]}}.

\bibitem{Song:2010mg}
H.~Song, S.~A. Bass, U.~Heinz, T.~Hirano, and C.~Shen, ``{200 A GeV Au+Au
  collisions serve a nearly perfect quark-gluon liquid},''
  \href{http://dx.doi.org/10.1103/PhysRevLett.106.192301,
  10.1103/PhysRevLett.109.139904}{{\em Phys. Rev. Lett.} {\bf 106} (2011)
  192301}, \href{http://arxiv.org/abs/1011.2783}{{\tt arXiv:1011.2783
  [nucl-th]}}.
[Erratum: Phys. Rev. Lett.109,139904(2012)].

\bibitem{Broniowski:2008vp}
W.~Broniowski, M.~Chojnacki, W.~Florkowski, and A.~Kisiel, ``{Uniform
  Description of Soft Observables in Heavy-Ion Collisions at s(NN)**(1/2) = 200
  GeV**2},'' \href{http://dx.doi.org/10.1103/PhysRevLett.101.022301}{{\em Phys.
  Rev. Lett.} {\bf 101} (2008)  022301},
\href{http://arxiv.org/abs/0801.4361}{{\tt arXiv:0801.4361 [nucl-th]}}.

\bibitem{Pratt:2015zsa}
S.~Pratt, E.~Sangaline, P.~Sorensen, and H.~Wang, ``{Constraining the Eq. of
  State of Super-Hadronic Matter from Heavy-Ion Collisions},''
  \href{http://dx.doi.org/10.1103/PhysRevLett.114.202301}{{\em Phys. Rev.
  Lett.} {\bf 114} (2015)  202301},
\href{http://arxiv.org/abs/1501.04042}{{\tt arXiv:1501.04042 [nucl-th]}}.

\bibitem{Bass:2017zyn}
S.~A. Bass, J.~E. Bernhard, and J.~S. Moreland, ``{Determination of
  Quark-Gluon-Plasma Parameters from a Global Bayesian Analysis},'' in {\em
  {26th International Conference on Ultrarelativistic Nucleus-Nucleus
  Collisions (Quark Matter 2017) Chicago,Illinois, USA, February 6-11, 2017}}.

\bibitem{Alqahtani:2017jwl}
M.~Alqahtani, M.~Nopoush, R.~Ryblewski, and M.~Strickland, ``{(3+1)D
  Quasiparticle Anisotropic Hydrodynamics for Ultrarelativistic Heavy-Ion
  Collisions},'' \href{http://dx.doi.org/10.1103/PhysRevLett.119.042301}{{\em
  Phys. Rev. Lett.} {\bf 119} (2017) no.~4, 042301},
\href{http://arxiv.org/abs/1703.05808}{{\tt arXiv:1703.05808 [nucl-th]}}.

\bibitem{Muller:1967zza}
I.~Muller, ``{Zum Paradoxon der Warmeleitungstheorie},''
\href{http://dx.doi.org/10.1007/BF01326412}{{\em Z. Phys.} {\bf 198} (1967)
  329--344}.

\bibitem{Israel:1976tn}
W.~Israel, ``{Nonstationary irreversible thermodynamics: A Causal relativistic
  theory},''
\href{http://dx.doi.org/10.1016/0003-4916(76)90064-6}{{\em Annals Phys.} {\bf
  100} (1976)  310--331}.

\bibitem{Israel:1979wp}
W.~Israel and J.~M. Stewart, ``{Transient relativistic thermodynamics and
  kinetic theory},''
\href{http://dx.doi.org/10.1016/0003-4916(79)90130-1}{{\em Annals Phys.} {\bf
  118} (1979)  341--372}.

\bibitem{Muronga:2001zk}
A.~Muronga, ``{Second order dissipative fluid dynamics for ultrarelativistic
  nuclear collisions},''
  \href{http://dx.doi.org/10.1103/PhysRevLett.88.062302}{{\em Phys. Rev. Lett.}
  {\bf 88} (2002)  062302}, \href{http://arxiv.org/abs/nucl-th/0104064}{{\tt
  arXiv:nucl-th/0104064 [nucl-th]}}.
[Erratum: Phys. Rev. Lett.89,159901(2002)].

\bibitem{Heinz:2005bw}
U.~W. Heinz, H.~Song, and A.~K. Chaudhuri, ``{Dissipative hydrodynamics for
  viscous relativistic fluids},''
  \href{http://dx.doi.org/10.1103/PhysRevC.73.034904}{{\em Phys. Rev.} {\bf
  C73} (2006)  034904},
\href{http://arxiv.org/abs/nucl-th/0510014}{{\tt arXiv:nucl-th/0510014
  [nucl-th]}}.

\bibitem{Bozek:2009dw}
P.~Bozek, ``{Bulk and shear viscosities of matter created in relativistic
  heavy-ion collisions},''
  \href{http://dx.doi.org/10.1103/PhysRevC.81.034909}{{\em Phys. Rev.} {\bf
  C81} (2010)  034909},
\href{http://arxiv.org/abs/0911.2397}{{\tt arXiv:0911.2397 [nucl-th]}}.

\bibitem{Shen:2014vra}
C.~Shen, Z.~Qiu, H.~Song, J.~Bernhard, S.~Bass, and U.~Heinz, ``{The
  iEBE-VISHNU code package for relativistic heavy-ion collisions},''
  \href{http://dx.doi.org/10.1016/j.cpc.2015.08.039}{{\em Comput. Phys.
  Commun.} {\bf 199} (2016)  61--85},
\href{http://arxiv.org/abs/1409.8164}{{\tt arXiv:1409.8164 [nucl-th]}}.

\bibitem{Denicol:2010xn}
G.~S. Denicol, T.~Koide, and D.~H. Rischke, ``{Dissipative relativistic fluid
  dynamics: a new way to derive the equations of motion from kinetic theory},''
  \href{http://dx.doi.org/10.1103/PhysRevLett.105.162501}{{\em Phys. Rev.
  Lett.} {\bf 105} (2010)  162501},
\href{http://arxiv.org/abs/1004.5013}{{\tt arXiv:1004.5013 [nucl-th]}}.

\bibitem{Denicol:2012cn}
G.~S. Denicol, H.~Niemi, E.~Molnar, and D.~H. Rischke, ``{Derivation of
  transient relativistic fluid dynamics from the Boltzmann equation},''
  \href{http://dx.doi.org/10.1103/PhysRevD.85.114047,
  10.1103/PhysRevD.91.039902}{{\em Phys. Rev.} {\bf D85} (2012)  114047},
  \href{http://arxiv.org/abs/1202.4551}{{\tt arXiv:1202.4551 [nucl-th]}}.
[Erratum: Phys. Rev.D91,no.3,039902(2015)].

\bibitem{Denicol:2014loa}
G.~S. Denicol, ``{Kinetic foundations of relativistic dissipative fluid
  dynamics},''
\href{http://dx.doi.org/10.1088/0954-3899/41/12/124004}{{\em J. Phys.} {\bf
  G41} (2014) no.~12, 124004}.

\bibitem{Baier:2007ix}
R.~Baier, P.~Romatschke, D.~T. Son, A.~O. Starinets, and M.~A. Stephanov,
  ``{Relativistic viscous hydrodynamics, conformal invariance, and
  holography},'' \href{http://dx.doi.org/10.1088/1126-6708/2008/04/100}{{\em
  JHEP} {\bf 04} (2008)  100},
\href{http://arxiv.org/abs/0712.2451}{{\tt arXiv:0712.2451 [hep-th]}}.

\bibitem{Florkowski:2010cf}
W.~Florkowski and R.~Ryblewski, ``{Highly-anisotropic and strongly-dissipative
  hydrodynamics for early stages of relativistic heavy-ion collisions},''
  \href{http://dx.doi.org/10.1103/PhysRevC.83.034907}{{\em Phys. Rev.} {\bf
  C83} (2011)  034907},
\href{http://arxiv.org/abs/1007.0130}{{\tt arXiv:1007.0130 [nucl-th]}}.

\bibitem{Martinez:2010sc}
M.~Martinez and M.~Strickland, ``{Dissipative Dynamics of Highly Anisotropic
  Systems},'' \href{http://dx.doi.org/10.1016/j.nuclphysa.2010.08.011}{{\em
  Nucl. Phys.} {\bf A848} (2010)  183--197},
\href{http://arxiv.org/abs/1007.0889}{{\tt arXiv:1007.0889 [nucl-th]}}.

\bibitem{Barz:1987pq}
H.~W. Barz, B.~Kampfer, B.~Lukacs, K.~Martinas, and G.~Wolf, ``{Deconfinement
  transition in anisotropic matter},''
\href{http://dx.doi.org/10.1016/0370-2693(87)90761-1}{{\em Phys. Lett.} {\bf
  B194} (1987)  15--19}.

\bibitem{Kampfer:1990qg}
B.~Kampfer, B.~Lukacs, G.~Wolf, and H.~W. Barz, ``{Description of the nuclear
  stopping process within anisotropic thermo hydrodynamics},''
\href{http://dx.doi.org/10.1016/0370-2693(90)91101-G}{{\em Phys. Lett.} {\bf
  B240} (1990)  297--300}.

\bibitem{Bazow:2013ifa}
D.~Bazow, U.~W. Heinz, and M.~Strickland, ``{Second-order (2+1)-dimensional
  anisotropic hydrodynamics},''
  \href{http://dx.doi.org/10.1103/PhysRevC.90.054910}{{\em Phys. Rev.} {\bf
  C90} (2014) no.~5, 054910},
\href{http://arxiv.org/abs/1311.6720}{{\tt arXiv:1311.6720 [nucl-th]}}.

\bibitem{Molnar:2016vvu}
E.~Molnar, H.~Niemi, and D.~H. Rischke, ``{Derivation of anisotropic
  dissipative fluid dynamics from the Boltzmann equation},''
  \href{http://dx.doi.org/10.1103/PhysRevD.93.114025}{{\em Phys. Rev.} {\bf
  D93} (2016) no.~11, 114025},
\href{http://arxiv.org/abs/1602.00573}{{\tt arXiv:1602.00573 [nucl-th]}}.

\bibitem{Molnar:2016gwq}
E.~Molnar, H.~Niemi, and D.~H. Rischke, ``{Closing the equations of motion of
  anisotropic fluid dynamics by a judicious choice of moment of the Boltzmann
  equation},''
\href{http://arxiv.org/abs/1606.09019}{{\tt arXiv:1606.09019 [nucl-th]}}.

\bibitem{Romatschke:2003ms}
P.~Romatschke and M.~Strickland, ``{Collective modes of an anisotropic quark
  gluon plasma},'' \href{http://dx.doi.org/10.1103/PhysRevD.68.036004}{{\em
  Phys. Rev.} {\bf D68} (2003)  036004},
\href{http://arxiv.org/abs/hep-ph/0304092}{{\tt arXiv:hep-ph/0304092
  [hep-ph]}}.

\bibitem{Tinti:2015xwa}
L.~Tinti, ``{Anisotropic matching principle for the hydrodynamics expansion},''
  \href{http://dx.doi.org/10.1103/PhysRevC.94.044902}{{\em Phys. Rev.} {\bf
  C94} (2016) no.~4, 044902},
\href{http://arxiv.org/abs/1506.07164}{{\tt arXiv:1506.07164 [hep-ph]}}.

\bibitem{Jaiswal:2013vta}
A.~Jaiswal, ``{Relativistic third-order dissipative fluid dynamics from kinetic
  theory},'' \href{http://dx.doi.org/10.1103/PhysRevC.88.021903}{{\em Phys.
  Rev.} {\bf C88} (2013)  021903},
\href{http://arxiv.org/abs/1305.3480}{{\tt arXiv:1305.3480 [nucl-th]}}.

\end{thebibliography}\endgroup


\providecommand{\href}[2]{#2}\begingroup\raggedright\endgroup
\end{document}